\documentclass[aps,pra,twocolumn]{revtex4-2}
\usepackage[latin9]{inputenc}
\usepackage{graphicx}
\usepackage{amssymb}
\usepackage{amsmath}
\usepackage{color, soul}
\usepackage{bm}
\usepackage[caption=false]{subfig}
\usepackage{placeins}
\usepackage{dsfont}
\usepackage{braket}

\newcommand{\w}{\omega}

\newcommand{\ntot}{n_{\rm tot}}
\newcommand{\epsk}{\epsilon_{\textbf{k}}}
\newcommand{\epsf}{\epsilon_f}

\newcommand{\epskf}{\epsilon_{\textbf{k}f}}

\newcommand{\ii}{{\rm i}}

\newcommand{\absval}[1]{\lvert #1 \rvert}

\RequirePackage[
   hyperindex,colorlinks,bookmarksnumbered,
   plainpages=true, pdfstartview=FitH,bookmarks=false
]{hyperref}
\hypersetup{linkcolor=blue,urlcolor=blue,citecolor=blue}

\definecolor{darkgreen}{rgb}{0,0.5,0}
\definecolor{darkblue}{rgb}{0,0,0.5}
\definecolor{purple}{rgb}{0.35,0,0.35}
\definecolor{orange}{rgb}{0.9,0.4,0}

\graphicspath{{./}{./figs/}}

\begin{document}
\title{
Kondo breakdown transitions and phase-separation tendencies \\
in valence-fluctuating heavy-fermion metals
}

\author{Pedro M. C\^onsoli}
\author{Matthias Vojta}
\affiliation{Institut f\"ur Theoretische Physik and W\"urzburg-Dresden Cluster of Excellence ct.qmat, Technische Universit\"at Dresden,
01062 Dresden, Germany}

\date{\today}

\begin{abstract}
The breakdown of the lattice Kondo effect in local-moment metals can lead to non-trivial forms of quantum criticality and a variety of non-Fermi-liquid phases. Given indications that Kondo-breakdown transitions involve criticality not only in the spin but also in the charge sector, we investigate the interplay of Kondo breakdown and strong valence fluctuations in generalized Anderson lattice models. We employ a parton mean-field theory to describe the transitions between deconfined fractionalized Fermi liquids and various confined phases. We find that rapid valence changes near Kondo breakdown can render the quantum transition first order. This leads to phase-separation tendencies which, upon inclusion of longer-range Coulomb interactions, will produce intrinsically inhomogeneous states near Kondo-breakdown transitions. We connect our findings to unsolved aspects of experimental data.
\end{abstract}

\maketitle


\section{Introduction}

Quantum criticality \cite{ssbook} in strongly correlated metals continues to be an exciting subject of condensed matter research \cite{stewart01,hvl07}. It involves a fascinating phenomenology including strange-metal behavior, strong-coupling critical points beyond the Landau-Ginzburg-Wilson paradigm, their interplay with non-trivial band topology, as well as instabilities to other phases such as unconventional superconductivity.
A particularly rich arena is that of multi-band system involving lattices of local moments. Local moments can be Kondo-screened, leading to heavy-fermion metallic behavior, or can induce various forms of symmetry-breaking or topological order. The breakdown of the Kondo effect has been theoretically argued to lead to non-trivial quantum phase transitions \cite{coleman01,si01} and possibly to topological non-Fermi-liquid states \cite{senthil03}. Experimentally, different signatures of Kondo breakdown have been identified in a number of compounds, such as CeCu$_{6-x}$Au$_x$ \cite{schroeder00}, YbRh$_2$Si$_2$ \cite{paschen04,friedemann10}, Ce$_3$Pd$_{20}$Si$_6$ \cite{custers12}, and CeCoIn$_5$ \cite{shishido05,analytis22}, but a comprehensive picture has not yet emerged.

\begin{figure}
	\centerline{
		\includegraphics[width=\columnwidth]{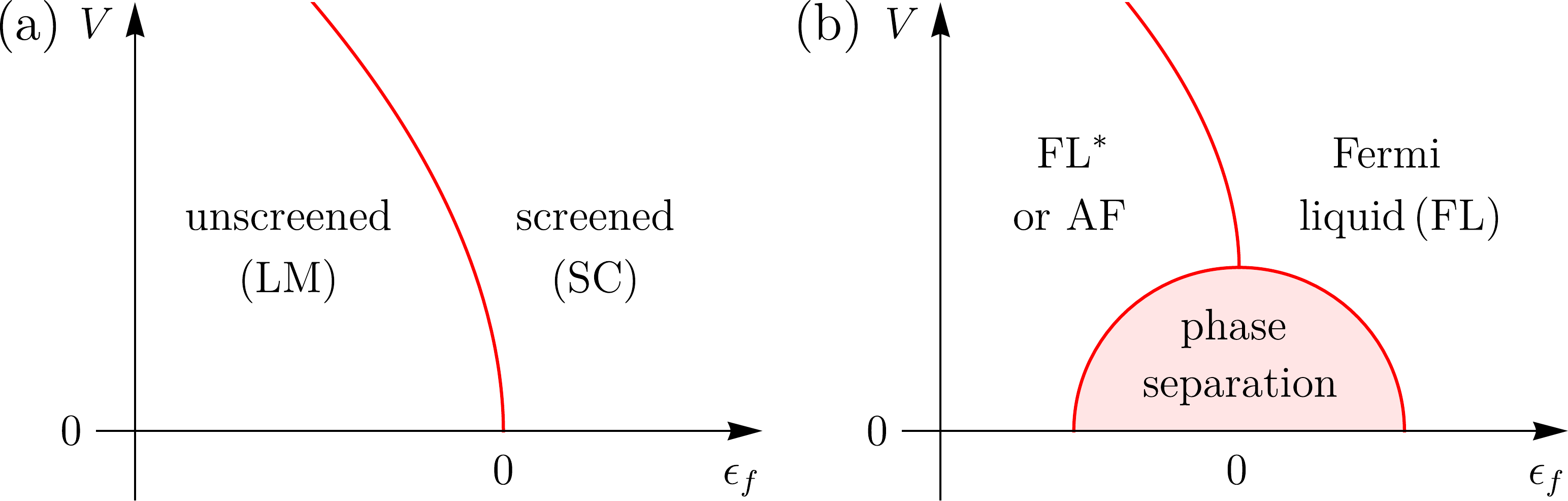}
	}
	\caption{
		(a) Schematic phase diagram of a single-impurity Anderson model with pseudogap density of states as function of $f$-level energy $\epsf$ and hybridization strength $V$, displaying a Kondo-breakdown transition between a screened (SC) and an unscreened (LM, local-moment) phase \cite{VF04,FV04}.
		(b) Phase diagram as before, but now for an Anderson-Heisenberg lattice model. The Kondo breakdown transition occurs between a heavy Fermi liquid (FL) and a light-electron phase which is either magnetically ordered (AF) or a fractionalized Fermi liquid (FL$^*$). In the valence-fluctuating region of small $V$, the transition is masked by phase-separation tendencies driven by finite inter-moment exchange interactions, leading to inhomogeneous states [this work].
	}
	\label{fig:pds}
\end{figure}

The breakdown of the Kondo effect has been traditionally assumed to give rise to a singular response in the spin sector but not necessarily in the charge sector \cite{coleman01,si01,senthil03,senthil04}, although the link to orbital-selective Mott transitions has been emphasized early on \cite{pepin08,mv10}. However, there is increasing evidence that this picture is incomplete.
This comes partly from pertinent experiments: Scaling behavior observed in the optical conductivity of YbRh$_2$Si$_2$ has been interpreted in terms of critical charge fluctuations \cite{paschen20}, and thermodynamic and transport measurements in CeRhIn$_5$ under pressure indicate the coincidence of magnetic and valence quantum critical points \cite{jaccard17}. In addition, an extended regime of non-Fermi liquid behavior has been found in YbAlB$_4$, which displays strong valence fluctuations \cite{matsumoto11,nakatsuji15}, and critical charge fluctuations have been observed in this compound using Mossbauer spectroscopy \cite{kobayashi22}.
These results have motivated corresponding theoretical activities: Critical charge fluctuations emerging at a Kondo-breakdown transition have recently been studied in a simplified Kondo-lattice model \cite{komijani19}. A different strand of work has linked various of the observed anomalies to quantum critical valence fluctuations \cite{miyake10,miyake12,miyake14}.
Remarkably, the link between Kondo breakdown and valence fluctuations can be made precise in simpler quantum impurity models, involving Kondo screening of isolated local moments. Here, Kondo breakdown can be induced, e.g., by a pseudogapped power-law density of states, $\rho(\w)\propto|\w|^s$, of the fermionic bath \cite{withoff90}, and the resultant quantum phase transition for bath exponents $s>1/2$ has been shown to involve spin and charge fluctuations on equal footing, with critical behavior for both \cite{VF04,FV04,pixley12,ingersent15}.
A similarly precise link for lattice models is missing, and thorough studies of the interplay and consequences of critical charge fluctuations and Kondo breakdown are scarce.

In this paper, we provide a step towards closing this gap. We study an Anderson lattice model in the limit of large Coulomb repulsion by suitable parton mean-field theories. We establish phase diagrams as function of both fixed total filling and fixed chemical potential, enabling us to investigate the interplay of Kondo breakdown and valence changes for a large range of model parameters. We show that this interplay is far more complex than for single-impurity models: A key new ingredient is that sizable valence changes across the transition imply a macroscopic redistribution of charge between bands which in turn strongly influences the electronic compressibility. As a result, the Kondo breakdown transition in the limit of small band hybridization is generically driven first order at fixed chemical potential, which implies phase separation at fixed density. We find that this phase-separation region extends to sizable values of hybridization, making it experimentally relevant. The inclusion of longer-range Coulomb interactions, which are not part of our modelling, will then generate inhomogeneous states by a mechanism of frustrated phase separation. In fact, inhomogeneous states whose origin has not been fully clarified have been observed in some correlated metals \cite{pfleiderer10,huxley15,montfrooij21}, and we argue that strong valence fluctuations are a possible cause for these phenomena.

The remainder of the paper is organized as follows:
In Sec.~\ref{sec:model} we introduce the Anderson lattice model which we decided to study and discuss various limits in parameter space. In particular, we give a general argument for the occurrence of phase separation near Kondo breakdown in the limit of small band hybridization.
Sec.~\ref{sec:parton} outlines the parton mean-field theory which we use to obtain concrete numerical results; these results are presented in Sec.~\ref{sec:mfres} and include a discussion of phases and phase diagrams obtained both in the canonical and grand-canonical settings.
Sec.~\ref{sec:ps} highlights the consequences of the phase separation which naturally emerges from the interplay of Kondo breakdown and valence fluctuations.
Finally Sec.~\ref{sec:fluct}, discussing physics not captured by the mean-field techniques used in this paper, argues that our qualitative findings are robust against fluctuation effects.
An outlook closes the paper. Technical details are relegated to appendices.


\section{Model and general considerations}
\label{sec:model}

\subsection{Anderson-Heisenberg lattice model} \label{subsec:Andlatt model}

The interplay of Kondo breakdown and valence fluctuations can be illustrated in a two-band Anderson lattice model, describing a strongly correlated $f$ band interacting with an uncorrelated $c$ band, which we supplement by an additional Heisenberg-type interaction between the $f$ sites. The Hamiltonian reads
\begin{align}
\mathcal{H} &=
-t\sum_{\langle ij \rangle,\sigma} c_{i\sigma}^\dagger c_{j\sigma}
+ \epsf \sum_{i,\sigma} f_{i\sigma}^\dagger f_{i\sigma} + U \sum_i n_{f,i\uparrow}
n_{f,i\downarrow} \nonumber\\
& +V \sum_{i,\sigma}(c_{i\sigma}^\dagger f_{i\sigma} + \mathrm{h.c.})
+ J \sum_{\langle ij \rangle} \left( \textbf{S}_i \cdot \textbf{S}_j - \frac{n_{f,i}n_{f,j}}{4} \right)
\label{eq:h}
\end{align}
in standard notation, with $n_{f,i\sigma}= f^\dagger_{i\sigma} f_{i\sigma}$ and $\textbf{S}_i = (1/2)\sum_{\sigma\sigma'} f^\dagger_{i\sigma} \boldsymbol{\tau}_{\sigma\sigma'} f_{i\sigma'}$, where $\boldsymbol{\tau} = \left(\tau_x, \tau_y, \tau_z\right)$ is a vector of Pauli matrices, and $\sigma=\uparrow,\downarrow$. For simplicity, the hybridization $V$ is assumed to be local, and both the $c$-electron hopping $t$ and the $f$-electron exchange $J$ are restricted to nearest-neighbor terms.
Here, we will primarily be interested in the limit of infinite on-site repulsion $U$, restricting the $f$ valence $n_f$ to fluctuate between $0$ and $1$. This is relevant, e.g., for Ce (Yb) compounds where local moments correspond to a $4f^1$ ($4f^{13}$) configuration, respectively, the latter requiring a particle-hole transformation.
In what follows we will specify band fillings according to $n_c = (1/N_s) \sum_{i\sigma} \langle c_{i\sigma}^\dagger c_{i\sigma}\rangle$, where $N_s$ is the number of lattice sites, such that a full conduction band corresponds to $n_c=2$ (and similarly for the $f$ band). We are mainly interested in the case $n_c\neq1$.

The Heisenberg coupling $J$ between the local moments in Eq.~\eqref{eq:h} can arise either from direct exchange or indirect Ruderman-Kittel-Kasuya-Yosida (RKKY) interactions. It competes with Kondo screening \cite{doniach77} and can drive the local-moment system into a magnetically ordered or a spin-liquid state. 
While the low-temperature state of the local moments depends on microscopic details, such as the lattice structure and the precise form of the Heisenberg coupling, we shall focus on cases where the $J$ term alone generates a quantum spin liquid (microscopically arising from some form of frustration). 
This enables a clear-cut definition of Kondo breakdown due to the absence of symmetry-breaking order: Upon increasing $J$, the heavy Fermi liquid (FL) phase of the model \eqref{eq:h} transitions into a metallic spin-liquid, dubbed fractionalized Fermi liquid (FL$^\ast$) \cite{senthil03,senthil04}, and these two phases can be sharply distinguished by the volume of their Fermi surfaces \cite{senthil03}.  As noted in the introduction, our goal is to study the fate of this FL$^\ast$--FL transition in the presence of strong valence fluctuations. In contrast, the onset of symmetry-breaking order causes Kondo and non-Kondo states to have the same Fermi volume, such that they can be adiabatically connected or separated by a Lifshitz-type transition \cite{senthil03,afkbnote}.

The density-density term in Eq.~\eqref{eq:h} has been added with an exchange mechanism in mind. If the $J$ interaction is instead dominantly of RKKY origin, then it will have long-range contributions and will moreover depend on the conduction-electron density. We will comment on these aspects towards the end of the paper.

We note that Kondo breakdown has been studied in an Anderson lattice model similar to Eq.~\eqref{eq:h} earlier in Ref.~\onlinecite{pepin08}, but with focus on the Kondo regime of weak valence fluctuations. We also note that inhomogeneous Kondo phases have appeared in Ref.~\onlinecite{zhu08}, but those are of very different character, i.e., weakly modulated and away from Kondo breakdown.

\subsection{Single impurity: Valence fluctuations vs. Kondo breakdown}

To set the stage, we discuss qualitative aspects of valence fluctuations and their interplay with Kondo breakdown in a single-impurity version of the Anderson model \eqref{eq:h}. We restrict ourselves to the low-temperature limit and place the conduction-band chemical potential at $\mu=0$ (not necessarily in the band center). We start by recalling the standard parameter regimes.
On the one hand, large positive $U$ and large negative $\epsf$ lead to a stable local moment, such that charge fluctuations are suppressed, $n_f\to1$, and a mapping to a Kondo model via a Schrieffer-Wolff transformation is justified.
On the other hand, if $\epsf$ is comparable to or smaller than the $c$-electron bandwidth, then $n_f$ strongly fluctuates, corresponding to a regime of intermediate valence.

Kondo screening corresponds to a situation where the coupling between the $f$ level and the $c$ band is relevant in the renormalization group sense, leading to effectively hybridized states. Thus, Kondo breakdown involves the loss of this hybridization. In the single-impurity case, a Kondo impurity is always screened as $T\to 0$ for a metallic host, but screening can break down if the conduction-band density of states $\rho(\w)$ vanishes at the Fermi level. The pseudogap case $\rho(\w)\propto|\w|^s$ \cite{withoff90,chen98,GBI98} leads to well studied continuous quantum phase transitions between an unscreened impurity at small $V$ (or small Kondo coupling) and a screened impurity at large $V$ (or large Kondo coupling),  Fig.~\ref{fig:pds}(a).

The traditional description of Kondo screening (or the breakdown thereof) is done in the Kondo limit. In a parton description, the onset of screening at $T=0$ involves the condensation of a charged slave boson which couples to a half-filled level of spinons \cite{coleman84,withoff90,senthil03}. However, a detailed analysis of the Kondo-breakdown quantum phase transitions in the single-impurity pseudogap Kondo model \cite{GBI98,VF04,FV04} has shown that, in this model, this picture is correct only for small bath exponents $s$. In contrast, for larger $s$ and in the presence of particle--hole asymmetry, the critical theory is of fermionic character \cite{VF04,FV04}, and the critical fixed point involves critical fluctuations in both spin and valence \cite{VF04,pixley12,ingersent15}. Importantly, the RG flow at Kondo-breakdown criticality is away from the Kondo limit and towards a regime of strong valence fluctuations with small effective $\epsf$ and $V$.
In the relevant pseudogap Anderson model with $U=\infty$, the transition can be traced all the way to the limit $V\to 0$, Fig.~\ref{fig:pds}(a): In this limit, $\epsf<0$ corresponds to an unscreened local moment and hence a Kondo-breakdown phase, whereas $\epsf>0$ represents an empty orbital (i.e. no local-moment degree of freedom), which is adiabatically connected and therefore equivalent to a Kondo-screened phase \cite{VF04,FV04}.

\subsection{Lattice: Kondo breakdown and phase separation}
\label{sec:genps}

\begin{figure*}
	\centerline{
		\includegraphics[width=\textwidth]{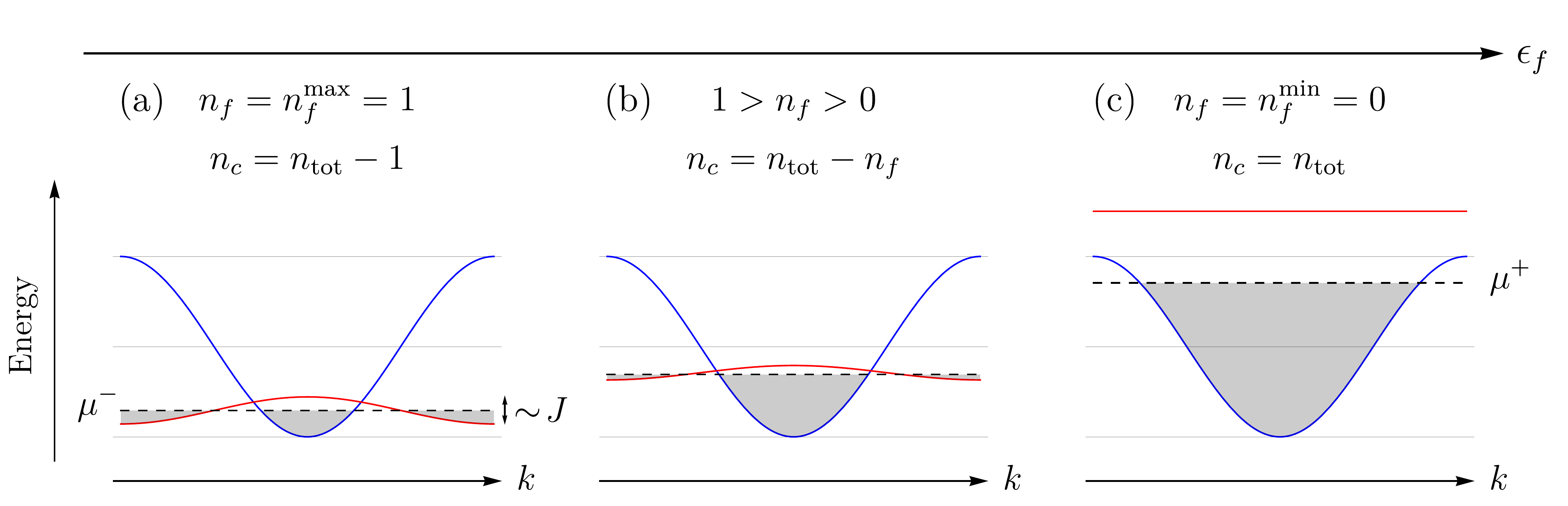}
	}
	\caption{
		Schematic evolution of the band dispersion and chemical potential, with $\epsf$ increasing from (a) to (c), illustrating the phase-separation tendency in the limit of small hybridization $V$. The blue and red curves represent unhybridized conduction electron and $f$-particle bands, respectively, for details see text.
	}
	\label{fig:schem}
\end{figure*}

Motivated by this insight, we now consider the $T=0$ physics of the Anderson lattice model \eqref{eq:h} upon varying $\epsf$ for small $V$. This is not the parameter regime traditionally discussed for heavy-fermion materials with well-established local moments, but opens the way to novel physics. As will become clear below, the chemical potential $\mu$ will play a key role when discussing the phases at finite total electron number.

The situation is particularly transparent in the limit $V\to 0$, followed by $J\to 0$. As above, $\epsf<\mu$ yields stable local moments, with $n_f\to1$. For non-vanishing $J$, these moments will either form a magnetically ordered state or realize a spin-liquid phase, depending on the amount of frustration or quantum fluctuations imposed by $J$ on the particular lattice, such that the resulting state is either a magnetically ordered metal or a fractionalized Fermi liquid (FL$^\ast$). In contrast, $\epsf>\mu$ results in a Fermi-liquid (FL) phase with an empty $f$ band, $n_f\to0$. Hence, tuning $(\epsf-\mu)$ through zero can drive an FL$^\ast$--FL transition. Importantly, this transition now necessarily coincides with a valence transition involving a massive change of the $f$-band occupation; such a change is not present for an FL$^\ast$--FL transition in the Kondo regime.

In a canonical perspective, i.e., for fixed $\ntot=n_c+n_f$, the change in $n_f$ across the transition must be compensated by a corresponding change in $n_c$. Assuming $1<\ntot<2$, this inevitably leads to phase separation near the transition, as we now show:
For $\epsf<\mu$ we have $n_c=\ntot-1$, and the chemical potential takes a value $\mu^-$ corresponding to this $c$-band filling, see Fig.~\ref{fig:schem}(a). Likewise, for $\epsf>\mu$ we have $n_c=\ntot$, and the chemical potential now takes the corresponding value $\mu^+$, Fig.~\ref{fig:schem}(c). This defines a range $\mu^-<\epsf<\mu^+$ where $\ntot-1<n_c<\ntot$ and therefore the $f$ band must be partially filled, Fig.~\ref{fig:schem}(b). In the considered limit $V,J\to 0$ where the $f$ band is flat, this implies a massively degenerate state. Now introducing a finite $J$ prefers spatial clustering of occupied $f$ sites in order to optimize magnetic exchange energy. In other words, phase separation into regions with filled and empty $f$ sites will occur. In a grand-canonical perspective with $\mu$ held fixed, this translates into a range of forbidden $\ntot$ upon variation of $\epsf$. We note that this argument does not rely on approximations or any further assumptions on the nature of the $f$-electron state, and hence applies even if the $f$-electron sector prefers magnetic order.
Furthermore, save for cases with flat $c$-bands, the argument does not depend on the precise form of the $c$-electron band structure and should hold irrespective of the details of the lattice geometry.
It also remains true whether or not the density-density interaction is included in Eq.~\eqref{eq:h}, and is unaffected by long-ranged RKKY interactions, because their spatial decay with distance still implies that most magnetic energy can be gained by spatial clustering of occupied $f$ sites.

This discussion clarifies that increasing $J$ amplifies phase separation tendencies. Conversely, increasing $V$ will diminish these tendencies, as finite $V$ leads to a dispersion of the effective $f$ band, such that hybridization-induced $f$-electron delocalization competes with their clustering due to $J$. The anticipated qualitative phase diagram is in Fig.~\ref{fig:pds}(b).


\section{Parton mean-field theory}
\label{sec:parton}

To obtain explicit results for the model \eqref{eq:h}, we employ a standard parton mean-field approach. As a consequence of taking the infinite-$U$ limit, the local $f$ Hilbert space is restricted to three states, the empty $\Ket{i,0}$ and the singly-occupied ones $\Ket{i,\sigma}$. The Hamiltonian thus reads
\begin{align}
\mathcal{H} - \mu \mathcal{N} &=
-t \sum_{\langle ij \rangle,\sigma} c_{i\sigma}^\dagger c_{j\sigma}
-\mu \sum_{i} \left( n_{c,i} + n_{f,i} \right)
\notag \\
& + V \sum_{i,\sigma} \left( \Ket{i,\sigma}\Bra{i,0} c_{i\sigma} + \mathrm{h.c.} \right)
+ \epsf \sum_{i} n_{f,i}
\notag \\
&+ J \sum_{\langle ij \rangle} \left( \textbf{S}_i \cdot \textbf{S}_j - \frac{n_{f,i}n_{f,j}}{4}\right)
\label{eq:HinfUAnd}
\end{align}
with $\mathcal{N}$ the total particle number operator, $n_{c,i} = \sum_{\sigma} c_{i\sigma}^\dagger c_{i\sigma}$, $n_{f,i} = \sum_{\sigma} \Ket{i,\sigma}\Bra{i,\sigma}$, and $\textbf{S}_i = \sum_{\sigma\sigma'} \Ket{i,\sigma}\Bra{i,\sigma'} \boldsymbol{\tau}_{\sigma\sigma'}/2$.
The constraint of no double occupancies is resolved using a slave-boson representation of the Hubbard operators \cite{coleman84}:
\begin{align}
\Ket{i,\sigma} \Bra{i,0} &= f_{i\sigma}^\dagger b_i, &
\Ket{i,0} \Bra{i,0} &= b_i^\dagger b_i,
\notag \\
\Ket{i,0} \Bra{i,\sigma} &=b_i^\dagger f_{i\sigma}, &
\Ket{i,\sigma} \Bra{i,\sigma'} &= f_{i\sigma}^\dagger f_{i\sigma'}.
\label{eq:slave boson rep}
\end{align}
Here, $b_i$ and $f_{i\sigma}$ are standard bosonic and fermionic operators, respectively. The representation \eqref{eq:slave boson rep} is faithful as long as one imposes the constraint $Q_i \!=\! b_i^\dagger b_i + \sum_{\sigma} f_{i\sigma}^\dagger f_{i\sigma} \!=\! \mathds{1}$ for every site $i$. With this, the Fock space at each site is reduced to the subspace spanned by the states $b_i^\dagger \Ket{0}$, $f_{i\uparrow}^\dagger \Ket{0}$, and $f_{i\downarrow}^\dagger \Ket{0}$, which are in a one-to-one correspondence with $\Ket{i,0}$, $\Ket{i,\uparrow}$, and $\Ket{i,\downarrow}$. Formally, the constraints are enforced by adding a term $\sum_{i} \lambda_{i} \left(Q_i-1\right)$,  where each $\lambda_{i}$ is a (fluctuating) Lagrange multiplier.

By inserting this representation into the Hamiltonian \eqref{eq:HinfUAnd}, the hybridization and Heisenberg terms are mapped onto three and four-operator contributions. Since these cannot be handled exactly, we resort to a mean-field approximation consisting of two steps. First, we replace the slave bosons $b_i$ ($b_i^\dagger$) by their expectation value $r_i$ ($r_i^*$), which is to be later determined by the minimization of a thermodynamic potential. Similarly, the $\lambda_{i}$ are replaced by static numbers. Second, we use the identity $\boldsymbol{\tau}_{\beta\beta'} \cdot \boldsymbol{\tau}_{\alpha\alpha'} = 2\delta_{\beta\alpha'}\delta_{\beta'\alpha} - \delta_{\beta\beta'}\delta_{\alpha\alpha'}$ to write
\begin{align}
\textbf{S}_i \cdot \textbf{S}_j
&= \frac{1}{4} \left( 2f_{i\alpha}^\dagger f_{i\alpha} - 2f_{i\beta}^\dagger f_{j\beta} f_{j\alpha}^\dagger f_{i\alpha}
- f_{i\beta}^\dagger f_{i\beta} f_{j\alpha}^\dagger f_{j\alpha} \right),
\label{eq:HeisenbergFierz}
\end{align}
where summations over repeated spins indices are implied, and decouple the quartic terms by pairing up bilinears with the same spin index. This amounts to decoupling the first and second quartic terms in the particle-hole and density-density channels, respectively. This decoupling aims at describing spin-liquid phases devoid of symmetry breaking. In particular, the particle-hole decoupling corresponds to the one which would be dictated within an SU($N$) large-$N$ limit; for discussions of alternative decoupling schemes we refer the reader the Appendix \ref{app:fsecmft}.

To proceed, we restrict our attention to a translation-invariant system with a mean-field unit cell composed of a single site. This allows us to replace the set of parameters $\left\{r_i, \lambda_i\right\}$ by two real variables, $r$ and $\lambda$, with nonzero $r$ signaling the presence of Kondo screening; furthermore, it implies that $n_{f,i} = \sum_{\sigma} \langle f_{i\sigma}^\dagger f_{i\sigma} \rangle$ is independent of the position $i$.
Besides that, we consider only mean-field solutions that preserve all spin and lattice symmetries, such that the hopping amplitude $\chi_{ij\sigma} = \langle f_{i\sigma}^\dagger f_{j\sigma} \rangle$ between nearest-neighbor sites $i$ and $j$ is spin- and bond-independent.
Finally, if we assume a Bravais lattice, the mean-field Hamiltonian takes the following Fourier-transformed form:
\begin{align}
\mathcal{H}_\mathrm{MF} - \mu\mathcal{N} &=
\sum_{\textbf{k}\sigma} \left[ \left(\epsk - \mu \right) c_{\textbf{k}\sigma}^\dagger c_{\textbf{k}\sigma} + \left(\epskf-\mu\right) f_{\textbf{k}\sigma}^\dagger f_{\textbf{k}\sigma} \right]
\notag \\ &
+ \widetilde{V} \sum_{\textbf{k}\sigma} \left( f_{\textbf{k}\sigma}^\dagger c_{\textbf{k}\sigma} + \mathrm{h.c.} \right)
+ N_s h_0,
\label{eq:HMF ph}
\end{align}
where $\widetilde{V} = rV$ is a renormalized hybridization and
\begin{align}
\epsk &= -t z \gamma_\textbf{k},
\\
\epskf &=
\epsf + \lambda + \frac{zJ}{4} \left(2r^2 - 1\right) - \chi z \gamma_\textbf{k},
\label{eq:epskf} \\
h_0 &= -\lambda \left(1-r^2\right) + \frac{zJ}{4} \left[
\left( \frac{\chi}{J} \right)^2 + \left(1-r^2\right)^2
\right].
\label{eq:h0}
\end{align}
$\chi$ is a mean-field parameter given by the self-consistency condition $\chi = (J/2) \sum_{\sigma} \langle f_{i\sigma}^\dagger f_{j\sigma}\rangle$, $z$ the coordination number, and $\gamma_\textbf{k} = (2/z) \sum_{\boldsymbol{\delta}} \cos \left(\textbf{k} \cdot \boldsymbol{\delta} \right)$ encodes the geometry of nearest-neighbor bonds, with being
$\boldsymbol{\delta}$ the set of primitive vectors of the underlying Bravais lattice.

One can diagonalize the quadratic mean-field Hamiltonian \eqref{eq:HMF ph} by means of a transformation
\begin{equation}
\begin{pmatrix}
c_{\textbf{k}\sigma} \\
f_{\textbf{k}\sigma}
\end{pmatrix}
=
\begin{pmatrix}
u_\textbf{k} & v_\textbf{k} \\
v_\textbf{k} & -u_\textbf{k}
\end{pmatrix}
\begin{pmatrix}
\alpha_{\textbf{k}\sigma+} \\
\alpha_{\textbf{k}\sigma-}
\end{pmatrix}
\end{equation}
with fermionic operators $\alpha_{\textbf{k}\sigma s}$ and coefficients $u_\textbf{k}$ and $v_\textbf{k}$ satisfying
\begin{align}
u_\textbf{k}^2 & = \frac{1}{2}\left( 1 + \frac{m_\textbf{k}}{\sqrt{m_\textbf{k}^2 + \widetilde{V}^2}} \right),
&
u_\textbf{k} v_\textbf{k} & = \frac{\widetilde{V}}{2\sqrt{m_\textbf{k}^2 + \widetilde{V}^2}} \; ,
\notag \\
v_\textbf{k}^2 & = \frac{1}{2}\left( 1 - \frac{m_\textbf{k}}{\sqrt{m_\textbf{k}^2 + \widetilde{V}^2}} \right),
&
m_\textbf{k} &= \frac{\epsk - \epskf}{2} \; .
\label{eq:diagonalization summed up}
\end{align}
The mean-field Hamiltonian thus becomes
\begin{equation}
\mathcal{H}_\mathrm{MF} - \mu\mathcal{N} = N_\mathrm{s}h_0 + \sum_{\textbf{k}s\sigma}
\left(E_{\textbf{k}s} - \mu\right) \alpha_{\textbf{k}\sigma s}^\dagger \alpha_{\textbf{k}\sigma s}
\end{equation}
with a dispersion given by
\begin{equation}
E_{\textbf{k}\pm} = \frac{\epsk + \epskf}{2} \pm \sqrt{m_\textbf{k}^2 + \widetilde{V}^2}.
\label{eq:ph dispersion}
\end{equation}

For computational simplicity, we will obtain explicit results for a two-dimensional square lattice, for which $z=4$ and $\boldsymbol{\delta} \in \left\{ \hat{\textbf{x}}, \hat{\textbf{y}} \right\}$ in units where the lattice constant is set to unity. We recall that our interest lies in spin-liquid phases realized in the local-moment sector of the model \eqref{eq:h}, and we note that the square-lattice spin-$1/2$ Heisenberg model realizes a spin-liquid phase if a second-neighbor coupling $J_2$ is included, with $0.45 < J_2/J < 0.56$ \cite{sandvik18,becca20,liu20}. The mean-field decoupling chosen here targets such a spin liquid, and we refrain from explicitly including $J_2$. We further note that the mean-field theory employed here displays additional solutions with larger mean-field unit cells, see Appendix~\ref{app:mfcell}, which we ignore for simplicity. Importantly, we expect that our qualitative conclusions are much more general; in particular, phase separation tendencies will also occur if the local-moment sector tends to magnetic order as opposed to spin-liquid behavior.


\section{Phases and phase diagrams from mean-field theory}
\label{sec:mfres}

We now discuss the numerical results obtained from the parton mean-field theory described above.

\subsection{Mean-field phases}
\label{subsec:mfphases}

We start by enumerating the possible phases; we recall that we restrict ourselves to situations where the mean-field parameters obey all lattice symmetries. The phases are primarily distinguished by whether or not the mean-field parameters $\chi$ and $r$ are non-zero.

First, there is a decoupled solution with $\chi=0$, $r=0$ that is realized at high temperatures. Physically, it reflects weakly interacting local moments, which only scatter conduction electrons weakly.

Second, there are solutions with $\chi\neq 0$, $r=0$ that correspond to fractionalized Fermi liquids (FL$^*$) \cite{senthil03,senthil04}. These are low-temperature phases without Kondo screening where only the $c$ (but not the $f$) electrons contribute to the volume of the Fermi surface, which is thus dubbed ``small'' and violates Luttinger's theorem. The fractionalization of the local moments becomes manifest as the $f$ sector of FL$^*$ corresponds to a U(1) spin liquid with a Fermi surface of neutral spinons. We note that the present mean-field theory displays additional solutions which, however, require larger mean-field unit cells, corresponding, e.g., to valence-bond solids or a $\pi$-flux spin liquid, and we will briefly comment on them in Appendix~\ref{app:mfcell}.

Third, there are solutions with $\chi\neq 0$, $r\neq 0$ where Kondo screening is active. Depending on whether the chemical potential is inside a band or in the band gap, these are either heavy Fermi liquids or Kondo insulators (KI), the latter displaying a total filling of $\ntot=2$ in the zero-temperature limit. The heavy Fermi liquids (FL$_{ns}$) can be further distinguished based on the number $n=1,2$ of Fermi sheets they possess and the sign $s=\pm$ of the mean-field parameter $\chi$ \cite{zhu08}. The inequivalence of states with the same $n$ but different $s$ can be established by dividing the square lattice under consideration (or, for the matter, any bipartite lattice) into two sublattices and applying a gauge transformation $f_{i\sigma} \to -f_{i\sigma}$ that only acts on sites $i$ belonging to one of them. This amounts to changing the sign of $\chi$ while additionally giving the slave-boson expectation values $r_i$ a staggered sign structure \cite{zhu08}.

We finally note that a phase with $\chi=0$, $r\neq 0$ does not exist, as non-zero $r$ inevitably generates non-zero $\chi$ because $f$ particles can hop via the $c$ band.

\begin{figure*}[t]
	\centerline{\includegraphics[width=\textwidth]{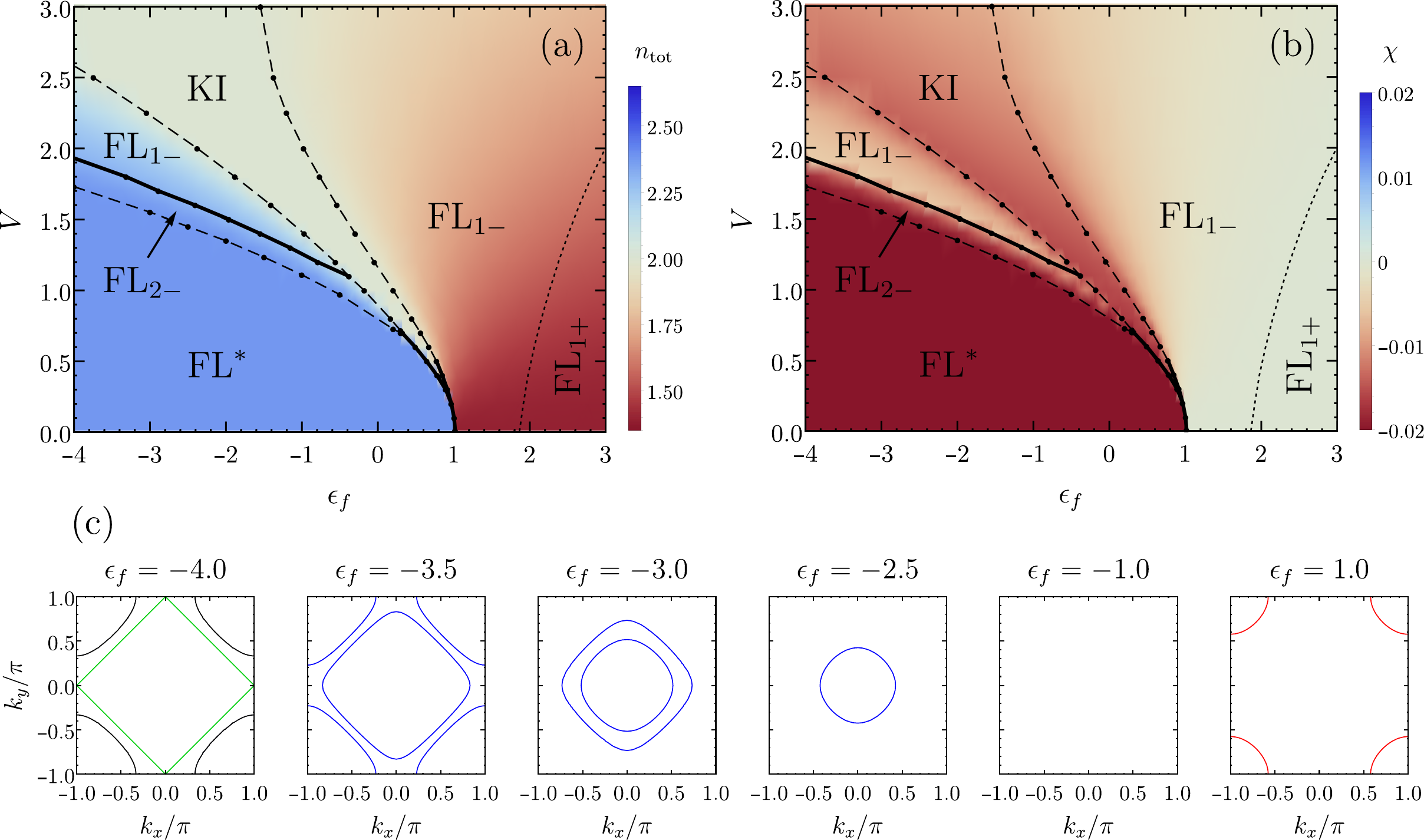}}
	\caption{(a,b) Mean-field phase diagram of the Anderson lattice model at a fixed chemical potential $\mu=1.0$ and with parameters $T=10^{-4}$ and $J=0.1$ in units of the $c$-electron hopping amplitude $t$. Dashed and solid lines indicate continuous and first-order transitions, respectively, whereas the color schemes show the variation of (a) the total filling $\ntot$ and (b) the mean-field parameter $\chi$. (c) Evolution of the Fermi surface for selected values of $\epsf$ and $V=1.7$. Red and blue lines correspond to the bands $E_{\textbf{k}-}=\mu$ and $E_{\textbf{k}+}=\mu$, respectively. In the leftmost panel, the spinon Fermi surface is represented in green and the small electron Fermi surface is shown in black.
	}
	\label{fig:ph gcpds}
\end{figure*}

\subsection{Grand-canonical phase diagram}

In the grand-canonical ensemble, i.e., at fixed $\mu$, the ground state of the system is obtained by minimizing the thermodynamic potential
\begin{equation}
\Omega = N_\mathrm{s} h_0
- 2k_\mathrm{B}T \sum_{\textbf{k}s}
\ln \left[ 1 + e^{-\beta \left( E_{\textbf{k}s} - \mu \right)} \right]
\end{equation}
with respect to the mean-field parameters $\left\{r, \lambda, \chi \right\}$. This yields the set of mean-field equations
\begin{align}
1 - r^2 & = n_f = \frac{2}{N_\mathrm{s}} \sum_{\textbf{k}} \left(u_\textbf{k}^2 n_{\textbf{k}-} + v_\textbf{k}^2 n_{\textbf{k}+} \right),
\label{eq:self-consistency D lambda}
\\
\lambda r  & = \frac{2V}{N_\mathrm{s}} \sum_{\textbf{k}} u_\textbf{k} v_\textbf{k} \left(n_{\textbf{k}-} - n_{\textbf{k}+} \right),
\label{eq:self-consistency D r}
\\
\chi & = \frac{J}{N_\mathrm{s}} \sum_{\textbf{k}} \gamma_\textbf{k} \left(u_\textbf{k}^2 n_{\textbf{k}-} + v_\textbf{k}^2 n_{\textbf{k}+} \right),
\label{eq:self-consistency D chi}
\end{align}
where $n_{\textbf{k}s} = \left[e^{\beta\left(E_{\textbf{k}s}-\mu\right)}+1\right]^{-1}$ is the Fermi-Dirac distribution function. We shall focus on the low-$T$ limit and perform the numerics at a small nonzero $T$ for numerical stability.

\begin{figure}
\centerline{\includegraphics[width=\columnwidth]{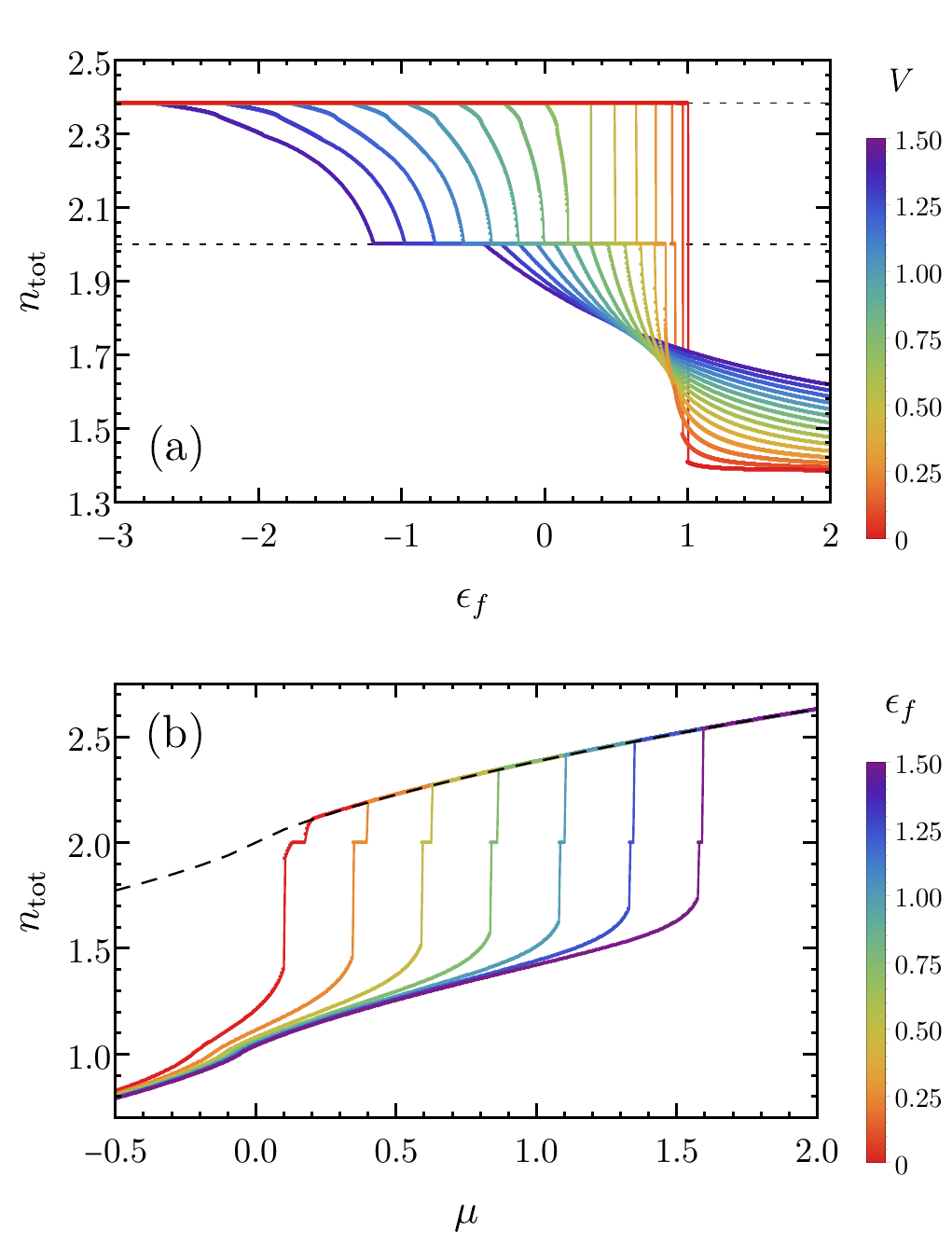}}
\caption{(a) Total filling, $\ntot$, as a function of $\epsf$ for various values of $V$ and with $T=10^{-4}$, $J=0.1$, and $\mu=1.0$. The two horizontal dashed lines correspond, in ascending order, to half-filling and to the maximum filling $\ntot^* \left(\mu=1\right) \approx 2.383$. (b) Also $\ntot$, but now as a function of $\mu$ at fixed $V=0.3$ and for different values of $\epsf$. The thick dashed line corresponds to the total filling in the absence of hybridization, when the $f$-band is half-filled, i.e., $\ntot^* \left(\mu\right) = 1 + n_{c}^* \left(\mu\right)$.
}
\label{fig:ph ntotXmu}
\end{figure}

Representative results are shown in Fig.~\ref{fig:ph gcpds}, which contains data for $T=10^{-4}$, $\mu=1$, and $J=0.1$ in units where the $c$-electron hopping amplitude $t=1$.
Panel (a) illustrates how $\ntot$ varies across a $V\,$--$\,\epsf$ phase diagram. Starting at fixed $V$, the limit $\epsf \to -\infty$ places us deep inside the FL$^{*}$ phase, where the $f$-band filling $n_f = 1$. As we increase $\epsf$ within the same phase, $n_f$ remains unchanged, but so does $n_c$, since the $c$ and $f$ bands are completely decoupled. Consequently, the total filling inside FL$^{*}$ is given by a constant $\ntot^* \left(\mu\right) = 1 + n_{c}^* \left(\mu\right)$.

Upon further increasing $\epsf$, the system is eventually driven out of the FL$^{*}$ phase as the screening of the local moments ensues. This process is signaled by the onset of a nonzero $r$, which, on account of Eq.~\eqref{eq:self-consistency D lambda}, reduces $n_f$ and thus causes the observed decrease in $\ntot$.
However, the nature of the phase transition and the resulting state depend on the hybridization strength. At large $V$, the system enters an FL$_{2-}$ phase through a continuous transition, across which $\ntot$ varies smoothly. On the other hand, once $V \lesssim 0.7$, a first-order transition occurs and leads to a KI or FL$_{1-}$ phase, depending on whether $V$ is moderate or low. In either case, the transition is accompanied by a jump in $\ntot$, which becomes increasingly pronounced as $V\to0$, Fig.~\ref{fig:ph ntotXmu}(a), as anticipated in Sec.~\ref{sec:genps}.

The phase diagram also exhibits various phase transitions that preserve the screening of the local moments, but entail changes in the topology of the Fermi surface. This is illustrated in Fig.~\ref{fig:ph gcpds}(c), which shows the evolution of the Fermi surface along the line $V=1.7$. The leftmost plot corresponds to the FL$^{*}$ phase, where one identifies a small electron Fermi surface centered at $\textbf{k}=\left(\pi,\pi\right)$ and a half-filled spinon Fermi surface. Once screening sets in, the latter becomes a Fermi surface of heavy electrons and loses its perfect nesting property. Upon further increasing $\epsf$, the two-sheet Fermi surface of  FL$_{2-}$ collapses onto a single-sheet Fermi surface of FL$_{1-}$ via a first-order Lifshitz transition which is accompanied by a jump in $\ntot$. Though not visible on the scale of Fig.~\ref{fig:ph ntotXmu}(a), this jump is present nonetheless and grows with $V$.
Further increase of $\epsf$ causes the Fermi surface to shrink, disappear, and finally reemerge as $\ntot$ falls below half-filling. This final transition from the KI to the FL$_{1-}$ state is discontinuous for $V\lesssim0.5$ and introduces yet another case in which the system experiences an abrupt large redistribution of charge, see Fig.~\ref{fig:ph ntotXmu}(a).

Let us now adopt a different perspective and analyze a few aspects of what happens as we vary $V$ at a fixed $\epsf$. In the interior of the FL$^{*}$ phase, this has no effect at all because the condition $r=0$ rescales the hybridization strength to zero. However, in the rest of the phase diagram, $\ntot$ follows a unified trend of approaching $2$ (half-filling) as $V$ overcomes all the other energy scales in the system. This behavior can be understood via Eq.~\eqref{eq:ph dispersion}, which indicates that, for any $r\ne0$, the two fermionic bands develop a gap proportional to $V$ in the limit $V\to\infty$. It is this increasing gap that stabilizes the KI at large $V$ and explains the change in curvature of its right boundary.

Fig.~\ref{fig:ph gcpds}(b) in turn illustrates how the mean-field parameter $\chi$ evolves across the phase diagram. A prominent feature therein is that solutions with $\chi<0$ are predominant, especially near the boundary of the FL$^*$ phase or, more specifically, when $r$ is small and $n_f$ is close to one.
This result was previously noted in Ref.~\cite{zhu08} and is related to the fact that $\chi>0$ generates $c$ and $f$ bands with the same the momentum dependence, whereas, for $\chi<0$, the minima of one band coincide with the maxima of the other. In the limit of small $r$, when the center of the $f$ band is close to $\mu$, the latter condition promotes the formation of a band gap, and hence produces a more stable solution by shifting the occupied states to lower energies.
On the other hand, our results also indicate that, as we increase $\epsf$ to a point where both $n_f$  and $\absval{\chi}$ approach zero, the existence of such a band gap ceases to be advantageous and the ground state acquires a positive $\chi$. This gives rise to a phase transition between FL$_{1-}$ and FL$_{1+}$ in the regime of positive $\epsf$.
The same density plot also reveals that $\chi$ jumps across first-order phase transitions in the model to accommodate the discontinuities in $\ntot$. In particular, it undergoes a sharp growth across the Lifshitz transition between FL$_{2-}$ and FL$_{1-}$, followed by an interval of non-monotonic behavior as the system enters the KI phase.

Finally, Fig.~\ref{fig:ph ntotXmu}(b) shows how $\ntot$ varies as a function of $\mu$ for different values of $\epsf$ and $V=0.3$. In preparation for an analysis in the canonical ensemble, we can consider the effect of tuning $\epsf$ at a fixed filling, such as $\ntot=1.5$. By starting from large values, where the ground state corresponds to a FL, and decreasing $\epsf$, we find that there is an extended interval preceding the FL$^{*}$ phase in which none of the previous homogeneous mean-field states realize the specified $\ntot$. This defines a forbidden range of $\mu$ and signals the phase separation tendency discussed in Sec.~\ref{sec:genps}. In the next section, we will confirm this connection by showing that the mean-field solution for such intermediate values of $\epsf$ is physically unstable.

Results for other values of $J$ and $0<\mu<zt$ are qualitatively similar; for $\mu<0$ the KI phase is not accessible.

\subsection{Canonical phase diagram}
\label{subsec:canpd}

If one approaches the problem from the perspective of the canonical ensemble, then a fourth self-consistency condition
\begin{equation}
\ntot = n_c + n_f =  \frac{2}{N_\mathrm{s}} \sum_{\textbf{k}} \left( n_{\textbf{k}-} + n_{\textbf{k}+} \right),
\label{eq:ntot}
\end{equation}
which fixes total filling to a specific value $\ntot$, must be added to the set of equations \eqref{eq:self-consistency D lambda} to \eqref{eq:self-consistency D chi}. The ground state of the system then corresponds to the mean-field solution that minimizes the free energy $F = \Omega + \mu \ntot N_s$.

\begin{figure}
\centerline{
\includegraphics[width=\columnwidth]{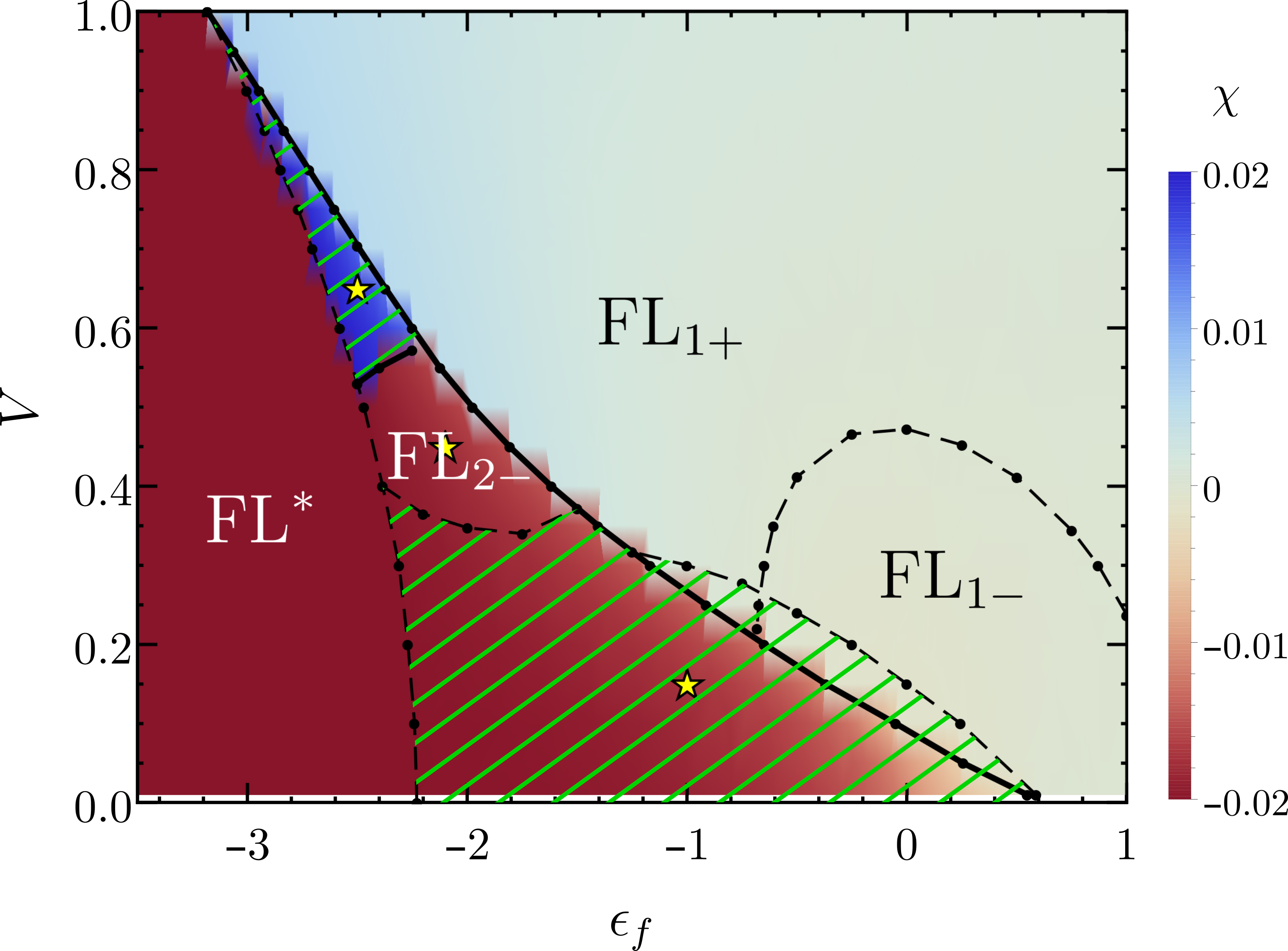}
}
\caption{Mean-field phase diagram of the Anderson lattice model at a fixed total filling of $\ntot=1.3$ and with parameters $T=10^{-4}$ and $J=0.1$ in units of the $c$-electron hopping amplitude $t$. The hatched regions indicate where the homogeneous mean-field solution with the lowest free energy has a negative compressibility $\kappa = \partial \ntot/ \partial \mu$. While we cannot access arbitrarily small $V$ for computational reasons, our results from Appendix~\ref{app:fsecmft} suggest that the area with $\kappa<0$ extends to $\epsf \to \infty$ in the limit $V,T\to0$. The actual regime of phase separation is larger than the hatched regions, see text and Appendix~\ref{app:maxwell}.
}
\label{fig:ph cpd}
\end{figure}

A representative phase diagram derived for $\ntot=1.3$, $J=0.1$, and at $T=10^{-4}$ is shown in Fig.~\ref{fig:ph cpd}. For any fixed $V$, we see that the limits of large negative and large positive $\epsf$ stabilize the FL$^{*}$ phase and an FL state, respectively, as in the grand canonical ensemble.
However, by repeating the calculation for each point in the phase diagram at a slightly larger filling ($\Delta\ntot = 10^{-3}\ntot$), we found that the intermediate regime now includes two regions in which the mean-field solutions have \textit{negative} charge compressibility $\kappa = \partial\ntot / \partial\mu$. In such regions, the concavity of $F\left(T,\ntot\right)$ around $\ntot=1.3$ implies that the system undergoes phase separation, since the configuration that minimizes the free energy is a mixture of two homogeneous states, with fillings above and below $\ntot=1.3$, respectively.

As predicted in Sec.~\ref{sec:genps}, one of the regions with $\kappa<0$ is connected to the limit $V\to0$, where the dispersion of $f$ electrons through the $c$ band is suppressed and the system profits from the formation of dense islands.
The second of such regions appears at larger $V$, where the physical grounds for phase separation are not as clear.
However, we recall that $\kappa>0$ is a necessary, but not sufficient, condition for the convexity of $F$. Therefore, phase separation is not restricted to occur in the hatched portions of the phase diagram. In Appendix~\ref{app:maxwell} we provide evidence that the model features a \textit{single} region of phase separation, which covers both hatched regions and therefore extends to sizable $V$.

\section{Beyond mean-field}
\label{sec:fluct}

In this section we quickly discuss physics beyond the zero-temperature mean-field analysis presented in Sec.~\ref{sec:mfres}. This discussion has two separate parts.

First, we discuss fluctuations effects on the FL--FL$^\ast$ transition, in part recalling the arguments of Refs.~\onlinecite{senthil03,senthil04}. As noted above, we have restricted our attention to mean-field solutions preserving both spin and translation symmetries. For FL$^\ast$ these correspond to homogeneous spin liquids in the $f$-electron sector. Determining whether such a phase is realized for a given microscopic spin model is a difficult task, with the list of numerically established cases growing \cite{lee08,savary17,knolle19}, although the precise nature of the liquid state has often not been clarified beyond doubt. Provided that the spin liquid exists as a stable phase of matter for the spin model under consideration, then the corresponding FL$^\ast$ phase exists as well \cite{senthil03}. In the case of a U(1) gauge group as discussed here, soft gauge fluctuations will produce singular contributions to the specific heat in both the FL$^\ast$ phase and the quantum critical regime of the FL--FL$^\ast$ transition \cite{senthil04}. In the latter, critical fluctuations of $r$ will lead to power-law spectral functions, i.e., critical Fermi surfaces \cite{senthil08}.

We note, however, that the quantum critical theory for lattice Kondo breakdown has not been established beyond doubt. In addition to the boson-condensation theory advocated in Refs.~\onlinecite{senthil03,senthil04,senthil08,pepin08} and used here, other proposals have been made \cite{si01,tsvelik16,komijani18,komijani19,chung20}. We recall that the single-site version of the boson-condensation theory does \emph{not} apply to the particle-hole-asymmetric pseudogap Kondo model where instead a fermionic theory is appropriate \cite{FV04}.
Exact numerical results have so far only been obtained for cases with Dirac fermions (instead of a full Fermi surface) \cite{hofmann19} and for one-dimensional models \cite{danu20}.
We hope that our work will also stimulate the construction of field theories explicitly taking into account valence fluctuations.

Second, we note that the ``global'' phase diagram as function of parameters $\epsf$ and $V$ may involve multiple different (ordered) states of local moments, in particular if their interactions are dominated by RKKY contributions, as those feature $k_F$ oscillations and are therefore density-dependent. Irrespective of such details, phase separation tendencies inevitably exist near the Kondo-breakdown transition.
We re-emphasize that the phase-separation tendencies also occur if the local-moment phase is an ordered magnet instead of a fractionalized Fermi liquid, see the discussion in Sec.~\ref{sec:genps}.


\section{Phase separation and inhomogeneous states}
\label{sec:ps}

The analysis so far has uncovered phase separation into regions of different electron density as a result of magnetic interactions. It can be expected that long-range Coulomb repulsion will render the phase-separated regime thermodynamically stable: The competition of short-range attractive and long-range repulsive forces leads to frustrated phase separation, and states with spatially inhomogeneous electron density will appear. Phenomena of this type have been discussed in the context of various strongly correlated systems \cite{emery93,zaanen98,vojta00,ortix06,maebashi12}, for instance for stripy charge order in underdoped cuprate superconductors.

Details of such inhomogeneous or modulated states will depend on both the precise form of the magnetic interactions and the lattice structure. We therefore leave a detailed microscopic study for future work and restrict ourselves to a few qualitative remarks.
(i) As the analysis in Appendix \ref{app:maxwell} shows, the system tends to separate into Kondo and non-Kondo regions, with different densities in both the $c$- and $f$-electron sectors. This is a key difference with respect to known examples of phase separation in Kondo lattice models, in which the constituents of the phase mixture, e.g. different itinerant magnetic states, have the same $f$-electron density \cite{pankratova21}.
(ii) While frustrated phase separation in the classical regime is expected to lead to islands or bubbles, quantum kinetic energy can lead to subdimensional extended structures, such as stripes in two-dimensional systems \cite{zaanen98,vojta00}.
(iii) The inevitable presence of quenched disorder in real crystals will lead to pinning phenomena of the resultant spatial structures, as disorder generically couples linearly to the particle density.


\section{Conclusions and outlook}
\label{sec:concl}

Motivated by recent experimental findings in local-moment metals, we have discussed Kondo-breakdown transitions in a generalized Anderson lattice model in the mixed-valence regime. We have argued that, away from the Kondo limit, Kondo breakdown is naturally accompanied by a sizable charge redistribution between different bands. Together with magnetic interactions between the local moments, this induces tendencies towards phase separation, which in turn leads to inhomogeneous states masking the Kondo-breakdown transition. We have obtained explicit results using a parton mean-field theory for a transition between FL and U(1) FL$^\ast$ phases, but we expect our qualitative results to be much more general. In particular, they will also apply if the non-Kondo state displays magnetic order.

To study this novel phenomenology in detail, future work should include numerically exact studies of suitable microscopic models. We note that the single-site dynamical mean-field approximation is not sufficient to capture the phase-separation tendency advocated here, as it cannot properly treat inter-moment interactions. In contrast, quantum Monte-Carlo studies of models designed to be sign-free \cite{hofmann19,danu22} appear to be a viable route.

Experimentally, emergent inhomogeneous states may be the reason for some of the puzzling phenomena observed in quantum critical heavy-fermion compounds. These include CeCu$_{6-x}$Au$_x$ \cite{schroeder00}, CeCoIn$_5$ \cite{shishido05,analytis22}, as well as YbAlB$_4$, which displays strong valence fluctuations \cite{matsumoto11,nakatsuji15}. An interesting case in point is CeRu$_2$Si$_2$, where recent thermodynamic measurements \cite{montfrooij21} found indications for the presence of magnetic clusters in a nominally clean compound. More microscopic measurements are clearly called for. Complementarily, new quantum chemistry studies could shed light on the necessary conditions to promote strong valence fluctuations and guide the synthesis of more candidates materials for this so-far largely unexplored regime of heavy-fermion metals.


\acknowledgments

We thank F. F. Assaad, B. Danu, B. Frank, and T. Grover for discussions and collaborations on related work.
Financial support from the Deutsche Forschungsgemeinschaft through SFB 1143 (project-id 247310070) and the W\"urzburg-Dresden Cluster of Excellence on Complexity and Topology in Quantum Matter -- \textit{ct.qmat} (EXC 2147, project-id 390858490) is gratefully acknowledged.


\appendix

\section{Saddle points of local-moment parton mean-field theories}
\label{app:mfcell}

As explained in the body of the paper, our focus is to describe the physics of Kondo-breakdown transitions. These are realized in a most clear-cut fashion for phase transitions between a heavy Fermi liquid and a fractionalized Fermi liquid \cite{senthil03,senthil04}. We therefore have restricted our attention to the simplest symmetric mean-field solutions for the non-Kondo phase, describing homogeneous spin liquids in the $f$-electron sector.

It is known, however, that the parton mean-field theories display other saddle points with larger unit cells. We quickly summarize the results obtained for the fermionic SU($N$) large-$N$ limit, where the decoupling is exclusively in the particle--hole channel with spinon hopping. Here, a solution with strong dimerization corresponding to a valence-bond solid has the lowest ground-state energy on most lattices. More homogeneous solutions can be stabilized by introducing either biquadratic \cite{affleck88,marston89} or ring exchange \cite{wang06}. On the square lattice, a sufficiently large biquadratic exchange stabilizes a homogeneous spin liquid which, however, has a $2\times1$ mean-field unit cell and realizes a Dirac spin liquid with spontaneous $\pi$-flux lattice \cite{affleck88,marston89}. We recall that, in parton mean-field theories, symmetries are realized projectively, hence such a solution does not break physical symmetries \cite{wen02}. Similar considerations apply to the Sp($2N$) large-$N$ where the decoupling is instead in the particle--particle channel \cite{read91,sachdev91}.

We emphasize again that our qualitative conclusions concerning phase separation and inhomogeneous states do not depend on details of the non-Kondo solutions, but only rely on the tendency of filled $f$-sites to cluster in the presence of $J$ when $V\to0$ and $n_f<1$, see Sec.~\ref{sec:genps}.


\section{Compressibility of homogeneous spin-liquid states in parton mean-field theories}
\label{app:fsecmft}

While the argument in favor of $f$-site clustering for a fractionally filled $f$ band is intuitive, it is not a priori clear that a given mean-field theory conforms to this physical expectation. In this appendix, we will present a quantitative analysis of different $f$-sector mean-field theories. Our starting point is the Hamiltonian
\begin{equation}
\mathcal{H}_{f} - \mu \mathcal{N}_{f} =
J \sum_{\langle ij \rangle} \left( \textbf{S}_i \cdot \textbf{S}_j - \frac{n_{f,i}n_{f,j}}{4}\right) -\mu \sum_{i} n_{f,i},
\label{eq:Hf}
\end{equation}
where $\epsf$ has been absorbed into $\mu$, as both quantities play the exact same role.

We first discuss the results for the decoupling scheme we considered in the main text. In this case, the relevant mean-field equations can be derived by setting $t=V=0$ in Eqs.~\eqref{eq:self-consistency D lambda}-\eqref{eq:ntot}. At fixed $n_f$, the slave-boson amplitude $r$ is determined directly from Eq.~\eqref{eq:self-consistency D lambda}, which reads $r^2 = 1-n_f$. Meanwhile, Eq.~\eqref{eq:self-consistency D r} becomes $\lambda r = 0$, implying that $\lambda=0$ for every $n_f<1$. The remaining mean-field equations are
\begin{align}
\chi &= \frac{J}{N_\mathrm{s}} \sum_{\textbf{k}} \frac{\gamma_{\textbf{k}}}{e^{\beta\left(\epskf - \mu\right)} + 1},
&
n_{f} &= \frac{2}{N_\mathrm{s}} \sum_{\textbf{k}} \frac{1}{e^{\beta\left(\epskf - \mu\right)} + 1},
\end{align}
with $\epskf = \lambda + zJ\left(1-2n_f\right)/4 - z\chi \gamma_{\textbf{k}}$.

\begin{figure}
	\centerline{\includegraphics[width=\columnwidth]{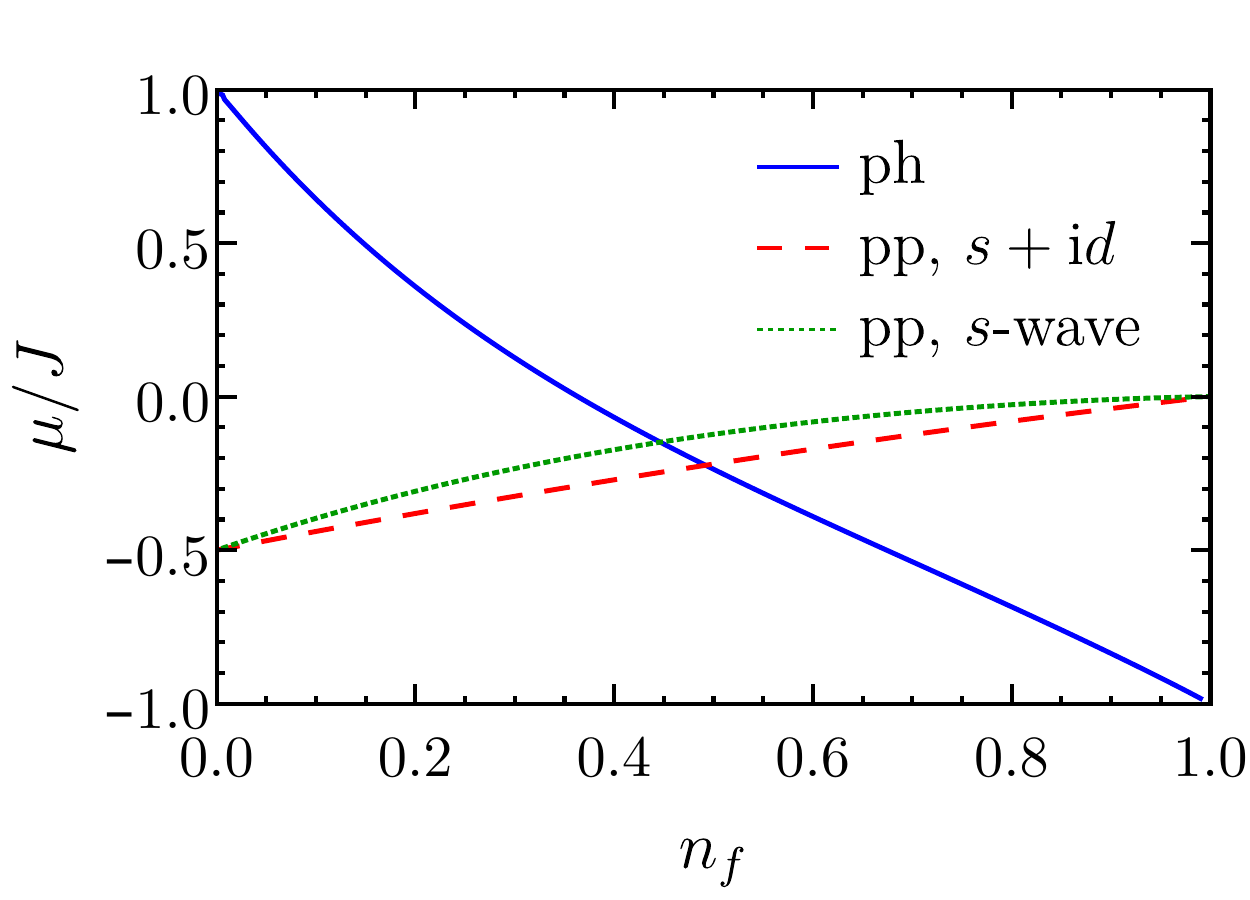}}
	\caption{Variation of the chemical potential $\mu$ with the $f$-band filling $n_f$ in the absence of hybridization, i.e., for the isolated $f$ sector of the infinite-$U$ Anderson-Heisenberg lattice model at $T/J=10^{-4}$. The blue solid curve corresponds to the particle--hole (ph) decoupling scheme presented in the main text, whereas the red dashed and green dotted lines follow from decoupling the interaction term in the particle--particle (pp) channel with $s$-wave and the energetically favored $s$+i$d$ pairing, respectively. Differently from the ph solution, for which $\kappa<0$, the pp solutions have positive compressibility.
	}
	\label{fig:fsector}
\end{figure}

In order to extract the compressibility $\kappa$, we solved the previous system of mean-field equations for various $n_f\in\left[0,1\right]$ and plotted $\mu$ as a function of $n_f$ in Fig.~\ref{fig:fsector}. There, one can verify that $\kappa<0$ over the entire range of fillings, confirming that the phase separation observed in the main text is indeed rooted in the limit of small $V$. In fact, this analysis also indicates that the hatched portion of the phase diagram in Fig.~\ref{fig:ph cpd} extends all the way to $\epsf\to\infty$ when $V\to0$. We have verified that this qualitative result is independent of whether the density--density term in Eq.~\eqref{eq:Hf} is included or not.

Next, we consider an alternative parton mean-field theory in which the interaction term in the Hamiltonian \eqref{eq:Hf} is decoupled in the particle--particle channel, formally obtained in the large-$N$ limit of Sp($2N$) \cite{read91,sachdev91,vojta00}. To perform the decoupling, we use the identity
\begin{equation}
\textbf{S}_{i}\cdot\textbf{S}_{j} - \frac{n_{f,i}n_{f,j}}{4}=
-\frac{1}{2} \left(f_{i\uparrow}^{\dagger}f_{j\downarrow}^{\dagger} \!-\! f_{i\downarrow}^{\dagger}f_{j\uparrow}^{\dagger}\right) \left(f_{j\downarrow}f_{i\uparrow} \!-\! f_{j\uparrow}f_{i\downarrow}\right)
\end{equation}
and define the mean-field averages
\begin{equation}
\frac{1}{2}\left\langle f_{j\downarrow}f_{i\uparrow}-f_{j\uparrow}f_{i\downarrow}\right\rangle =
\begin{cases}
\eta_{x} & \text{if }\left(ij\right)\text{ is horizontal}\\
\eta_{y} & \text{if }\left(ij\right)\text{ is vertical}
\end{cases}
\end{equation}
for nearest-neighbor sites $i$ and $j$. By doing so, we allow pairing fields to break the discrete rotational symmetry of the lattice, but require full translational symmetry with a single site per unit cell. One obtains the mean-field Hamiltonian
\begin{equation}
\mathcal{H}_{f,\rm{MF}} - \mu N_{f} = N_{\rm{s}}h_0 + \sum_{\textbf{k}} \Psi_\textbf{k}^\dagger \mathbb{M}_\textbf{k} \Psi_\textbf{k}
\label{eq:HMFpp1}
\end{equation}
with
\begin{align}
\mathbb{M}_{\textbf{k}} &=
\begin{pmatrix}
\lambda-\mu & -\Delta_{\textbf{k}} \\
-\Delta_{\textbf{k}}^{*} & -\left(\lambda-\mu\right)
\end{pmatrix},
&
\Psi_{\textbf{k}} &=
\begin{pmatrix}
f_{\textbf{k}\uparrow} \\
f_{-\textbf{k}\downarrow}^{\dagger}
\end{pmatrix},
\end{align}
and
\begin{align}
\Delta_{\textbf{k}} &= 2J\left(\eta_{x}\cos k_{x}+\eta_{y}\cos k_{y}\right), \notag \\
h_{0} &= -\mu +\lambda r^{2}+2J\left(\left|\eta_{x}\right|^{2}+\left|\eta_{y}\right|^{2}\right).
\label{eq:pp h0}
\end{align}
This free-fermion problem can be diagonalized by a standard Bogoliubov transformation, after which
\begin{equation}
\mathcal{H}_{f,\rm{MF}} - \mu N_{f} = N_{\rm{s}}h_0 -\sum_{\textbf{k}} E_\textbf{k} + \sum_{\textbf{k}\sigma} E_\textbf{k} \gamma_{\textbf{k}\sigma}^\dagger \gamma_{\textbf{k}\sigma}
\label{eq:HMFpp2}
\end{equation}
with a dispersion $E_\textbf{k} = \sqrt{\left(\lambda-\mu\right)^2 + \absval{\Delta_\textbf{k}}^2}$. The grand-canonical potential is then given by
\begin{equation}
\Omega = N_{s}h_{0}
- 2k_{{\rm B}}T \sum_{\textbf{k}} \ln \cosh \left( \frac{\beta E_{\textbf{k}}}{2} \right)
\label{eq:Omegapp}
\end{equation}
apart from irrelevant constants. By minimizing Eq.~\eqref{eq:Omegapp} with respect to the complex parameters $\left\{\eta_{x}^*,\eta_{y}^*\right\}$, we find
\begin{equation}
\eta_{\gamma} = \frac{1}{2N_{s}} \sum_{\textbf{k}} \tanh \left( \frac{\beta E_\textbf{k}}{2} \right) \frac{\Delta_\textbf{k} \cos k_\gamma}{E_\textbf{k}}
\label{eq:ppMFeqs}
\end{equation}
for $\gamma = x,y$. In principle, this yields a set of four mean-field equations, but, since the complex phases of $\eta_{x}$ and $\eta_{y}$ do not enter the energetics of the system independently, we can eliminate one of them by fixing a gauge (e.g., by taking $\mathrm{Im} \eta_{x} = 0$). Moreover, the condition $N_{\rm{s}} n_f = -\partial \Omega/\partial \mu$ fixes the chemical potential via
\begin{equation}
n_f = 1 + \frac{\mu-\lambda}{N_{s}} \sum_{\textbf{k}} \frac{1}{E_\textbf{k}} \tanh \left( \frac{\beta E_\textbf{k}}{2} \right).
\label{eq:ppMFnf}
\end{equation}

By solving the system of Eqs.~\eqref{eq:ppMFeqs} and \eqref{eq:ppMFnf}, we find that states with $s+\ii d$ pairing, for which $\eta_{y}=\ii \eta_{x}$, are favored \cite{kotliar88,sachdev91}. The corresponding change in $\mu$ as a function $n_f$ is plotted in Fig.~\ref{fig:fsector} and, remarkably, we see that $\kappa>0$ over the entire range of $n_f$. The same conclusion holds for higher-energy mean-field solutions with different pairing symmetries, such as $s$-wave pairing. Based on this observation, we conclude that the mean-field theory constructed from a pure particle-particle decoupling is completely oblivious to the phase separation tendencies of the model, and therefore becomes problematic away from half-filling.


\section{Maxwell construction for phase separation in the canonical ensemble}
\label{app:maxwell}

\begin{figure*}
	\centerline{
		\includegraphics[width=\textwidth]{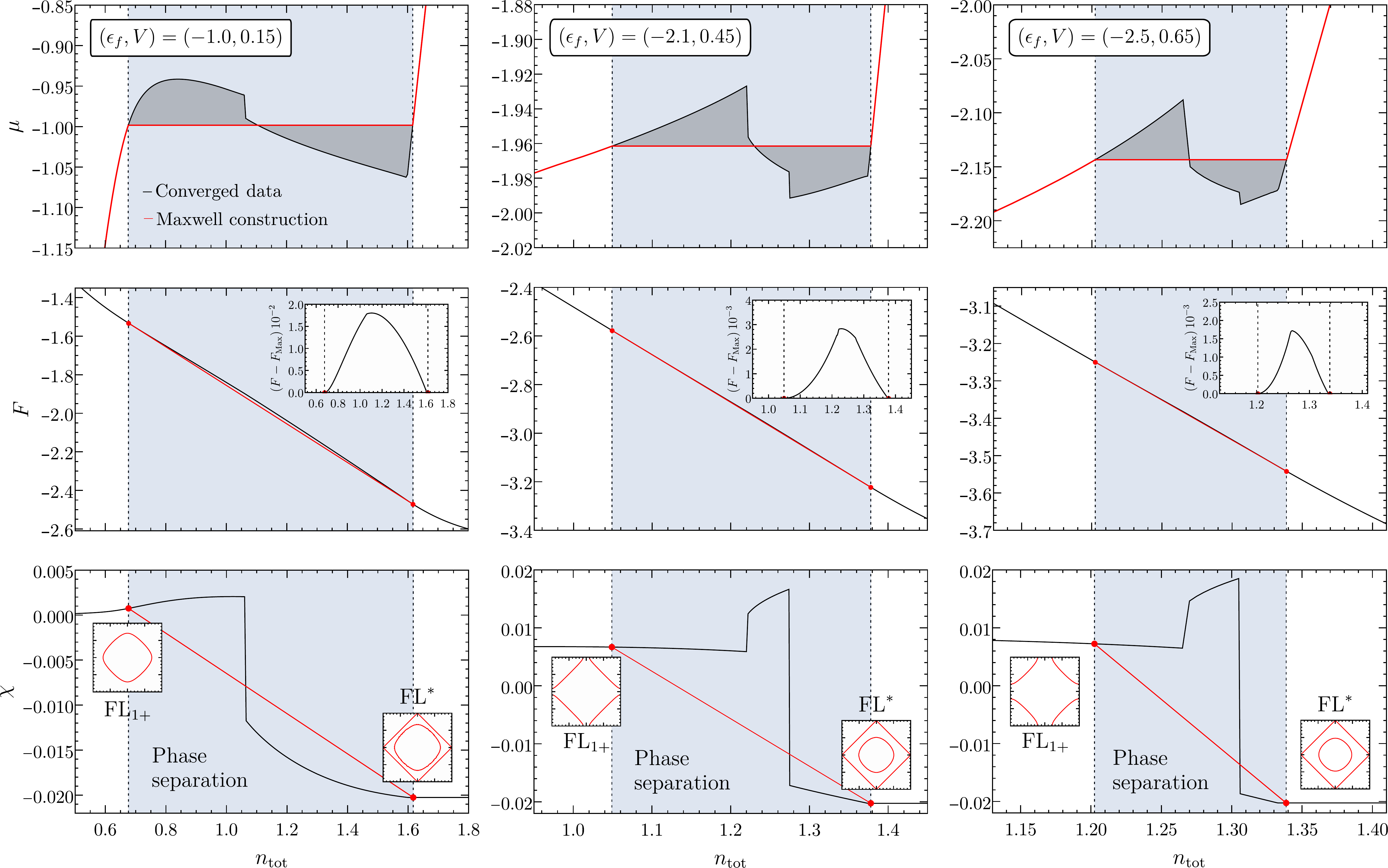}
	}
	\caption{Maxwell constructions and characterization of the phase mixtures at the three different points highlighted in the phase diagram of Fig.~\ref{fig:ph cpd}. The top row superimposes the mean-field solutions (black) for the chemical potential $\mu$ at different $n_{\mathrm{tot}}$ with the curve determined by the Maxwell construction (red). The shaded region delimited by the vertical gridlines indicates the range of fillings in which the system undergoes phase separation. The middle row depicts the evolution of the free energy $F$ in the same interval of $\ntot$. Black and red curves correspond, as before, to the homogeneous mean-field solution and phase mixture, respectively, and the difference between the two is shown in a smaller scale in the insets. Finally, the bottom row illustrates the variation of the mean-field parameter $\chi$. Together with the Fermi surfaces shown in the insets, this allows the identification of the two states that are mixed in the region of phase separation.
	}
	\label{fig:maxwell}
\end{figure*}

In this Appendix, we employ the Maxwell construction for different points in the canonical phase diagram of Fig.~\ref{fig:ph cpd} to investigate the occurrence phase separation in more detail. As we will show below, this provides deeper insight into the physical origin of the phenomenon in the model.

In Sec.~\ref{subsec:canpd}, we discussed that the existence of mean-field solutions with negative charge compressibility $\kappa$ is a sufficient, but not a necessary, condition for phase separation to take place. The key ingredient behind phase separation is rather that the free energy $F\left(T,\ntot\right)$ built from an ensemble of homogeneous states (as the one we considered in Sec.~\ref{subsec:canpd}) becomes a concave function of $\ntot$. Indeed, whenever this happens, one can lower the free energy at a given value of $\ntot$ by forming a mixture of two states with fillings $\ntot^{\left(1\right)}<\ntot$ and $\ntot^{\left(2\right)}>\ntot$, which appear in proportions such that the average filling is fixed to $\ntot$. According to the well-known Maxwell construction \cite{callen1985}, the values of $\ntot^{\left(1\right)}$ and $\ntot^{\left(2\right)}$ are determined by simultaneously fulfilling the conditions
\begin{align}
F\left( T,\ntot^{\left(1\right)} \right) &= F\left( T,\ntot^{\left(2\right)} \right), \\
\frac{\partial F}{\partial \ntot} \left( T,\ntot^{\left(1\right)} \right) &= \frac{\partial F}{\partial \ntot} \left( T,\ntot^{\left(2\right)} \right).
\end{align}

We carried out this procedure for the three different points marked by stars in the phase diagram of Fig.~\ref{fig:ph cpd}, one in each of the disjoint regions of negative $\kappa$ plus a third point in between. Figure \ref{fig:maxwell} shows how the behavior of the chemical potential $\mu$, the free energy $F$, and the mean-field parameter $\chi$ as a function of $\ntot$ changes from before (black) to after (red) the Maxwell construction.
In analyzing these results, two main features stand out. First, \emph{all three} $\left(\epsf,V\right)$ points give rise to phase separation at the filling $\ntot=1.3$ adopted in Fig.~\ref{fig:ph cpd}. We emphasize that this applies even to the intermediate-$V$ point, which has a homogeneous mean-field solution with positive $\kappa$. Second, the bottom row of plots in Fig.~\ref{fig:maxwell} reveals that the two constituents of the phase mixture are, in all three cases, FL$_{1+}$ and FL$^{*}$ states.
Put together, these observations strongly suggest that the separate domains with negative $\kappa$ shown in Fig.~\ref{fig:ph cpd} belong to the \emph{same} region of phase separation, which emanates from $V=0$ and extends up to $V\approx1$. If this is indeed true, then there is a \emph{single} physical mechanism driving phase separation in the model, namely that the gain in magnetic energy at small $V$ outweighs the delocalizing tendencies due to hybridization between the $c$ and $f$ bands.

In addition to the points raised above, we also note that the difference in the total filling $\Delta \ntot = \ntot^{\left(2\right)} - \ntot^{\left(1\right)}$ of the two homogeneous states that are combined in the phase mixture changes substantially as a function of $V$. More concretely, it increases from $\Delta\ntot \approx 0.14$ at the point with largest $V$ (right column) to $\Delta\ntot \approx0.94$ at the point with the smallest $V$ (left column).


\end{document}